\documentclass[letter]{aa} 
\usepackage{graphicx}
\usepackage{lscape}
\usepackage{txfonts}
\usepackage{natbib}
\usepackage{sidecap}
\usepackage{comment}
\usepackage{color}
\usepackage{booktabs}
\usepackage{array}

\begin{document}
\title{Massive open star clusters using the VVV survey}
\subtitle{III. A young massive cluster at the far edge of the Galactic bar\thanks{Based on observations taken within the ESO VISTA Public Survey VVV 
(programme ID 179.B-2002), and with ISAAC, VLT, ESO (programme 087.D-0341A).}}
\author{S. Ram\'irez Alegr\'ia\inst{1,2} \and J. Borissova\inst{1,2} \and A.N. Chen\'e\inst{3} \and E. O'Leary\inst{3} \and P. Amigo\inst{1,2}
\and D. Minniti\inst{2,4} \and R. K. Saito\inst{5} \and D. Geisler\inst{6} \and R. Kurtev\inst{1,2}\and M. Hempel\inst{2,4} \and M. Gromadzki
\inst{1} \and J. R. A. Clarke\inst{1} \and I. Negueruela\inst{7} \and A. Marco\inst{7} \and C. Fierro\inst{1,8} \and C. Bonatto\inst{9} \and M. Catelan
\inst{2,4} }

\institute{Instituto de F\'isica y Astronom\'ia, Universidad de Valpara\'iso, Av. Gran Breta\~na 1111, Playa Ancha, Casilla 
5030, Valpara\'iso, Chile \email{sebastian.ramirez@uv.cl} 
\and The Millennium Institute of Astrophysics (MAS), Santiago, Chile
\and Gemini North Observatory, USA
\and Pontificia Universidad Cat\'olica de Chile, Instituto de Astrof\'isica, Av. Vicu\~na Mackenna 4860, 782-0436 Macul, Santiago, Chile 
\and Universidade Federal de Sergipe, Departamento de F\'isica, Av. Marechal Rondon s/n, 49100-000, S\~ao Crist\'ov\~ao, SE, Brazil
\and Departamento de Astronom\'ia, Casilla 160-C, Universidad de Concepci\'on, Chile
\and Departamento de F\'isica, Ingenier\'ia de Sistemas y Teor\'ia de la Se\~nal, Universidad de Alicante, Spain
\and Escuela Superior de F\'isica y Matem\'aticas del Instituto Polit\'ecnico Nacional, Unidad Profesional Adolfo L\'opez Mateos, M\'exico
\and Universidade Federal do Rio Grande do Sul, Departamento de Astronomia, CP 15051, RS, 91501-970 Porto Alegre, Brazil
\\} 

\abstract
{Young massive clusters are key to map the Milky Way's structure, and near-IR large area sky surveys have contributed strongly to the discovery of new 
obscured massive stellar clusters.}
{We present the third article in a series of papers focused on young and massive clusters discovered in the VVV survey. This article is dedicated to the 
physical characterization of VVV\,CL086, using part of its OB-stellar population.}
{We physically characterized the cluster using $JHK_S$ near-infrared photometry from ESO public survey VVV images, using the VVV-SkZ pipeline, and 
near-infrared $K$-band spectroscopy, following the methodology presented in the first article of the series.}
{Individual distances for two observed stars indicate that the cluster is located at the far edge of the Galactic bar. These stars, which are probable cluster 
members from the statistically field-star decontaminated CMD, have spectral types between O9 and B0\,V. According to our analysis, this young cluster 
($1.0$ Myr $<$ age $< 5.0$ Myr) is located at a distance of $11^{+5}_{-6}$ kpc, and we estimate a lower limit for the cluster total mass of 
$(2.8^{+1.6}_{-1.4})\cdot10^3 {M}_{\odot}$. It is likely that the cluster contains even earlier and more massive stars.}
{}
 \titlerunning{VVV open star clusters III}
\maketitle


\section{Introduction}\label{intro}

Young massive clusters (cluster mass $M>10^3  M_{\odot}$, \citealt{hanson07}) are fundamental pieces for the study of Galactic structure. Because of 
their youth, they give information related to the recent Galactic massive stellar formation history. They are also excellent tracers of star formation regions.

 In the Milky Way, we find regions with intense stellar formation activity in the Galactic centre, where the Arches \citep{nagata93,figer02}, Center 
\citep{krabbe95,paumard06,figer08}, and the Quintuplet \citep{glass90,okuda90,nagata90,figer99} clusters are located; in the Carina-Sagittarius arm, which 
hosts Westerlund 2 \citep{westerlund61,rauw07,ascenso07}, Trumpler 14 \citep{ascenso07b,sanchawala07,sana10}, and NGC 3603 \citep{goss69,harayama08}; 
and the close edge of the Galactic bar. In the last region, several clusters with a massive population of red supergiant (RSG) stars are found: RSGC1 
\citep{figer06,davies08}, RSGC2 \citep{davies07}, RSGC3 \citep{alexander09,clark09a}, Alicante\,7 \citep{negueruela10}, Alicante\,8 \citep{negueruela11}, 
and Alicante\,10 \citep{gonzalezfernandez12}, but also younger clusters with a mixed population of OB-type and RSG stars (Masgomas-1, \citealt{ramirezalegria12}).
By symmetry, we expect massive star clusters at the far edge of the Galactic bar. Until now, only one massive cluster is known in this region: Mercer 81 
\citep{davies12}.
 
  Owing to the distance and extinction expected for stars located at the far edge of the bar, clusters in that region
 must be observed with near-infrared filters, which are less affected by interstellar extinction, and with a high spatial resolution, to be able to resolve
 the distant cluster population. One of the most recent near-IR surveys, the ESO public survey VISTA Variables in the V\'ia L\'actea (VVV, 
 \citealt{minnitiVVV10,saito10,saito12}), is a perfect tool for this exploration, covering the Galactic bulge and the adjacent disk 
 region, including the far edge of the bar with a spatial resolution of 0.34 arcsec pix$^{-1}$ in \textit{ZYJH$K_S$} filters.
  
 We present infrared spectrophotometric observations for VVV\,CL086 \citep{borissova11}, a massive cluster found in the direction of the Perseus arm 
($l=340\fdg001$, $b=-0\fdg293$), similar to that of Mercer 81 ($l=338\fdg400$, $b=+0\fdg100$). Data observation and reduction are described in Section 
\ref{observations}, near-IR photometry (CMD) and spectroscopy (spectral classification) are presented in Section \ref{resultados}. Cluster physical 
characterization is the topic of Section \ref{discusion}, and the general conclusions are given in the final Section \ref{conclusiones}.
 
\begin{figure}
\centering
\includegraphics[width=7.1cm,angle=0]{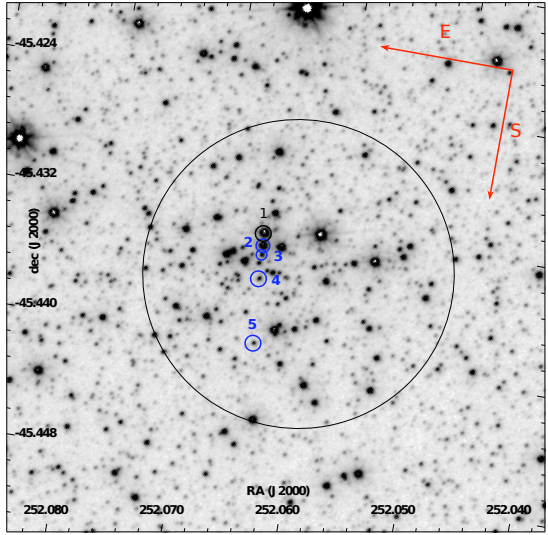}
       \caption{VVV $K_S$ image for VVV\,CL086. The image is 2$\times$2 arcmin$^{2}$ size and the cluster radius of 35$\arcsec$ 
       \citep{borissova11} is shown with a black circle. Small blue (OB-type stars) and black circles mark the position of the spectroscopically 
       observed stars.}
       \label{vvvCL086_Ks}
\end{figure}
 

\section{Observations}\label{observations}

The photometric data used in this work are part of the ESO Public Survey VVV, which observes with the VIRCAM at the VISTA
telescope. VIRCAM has a 16-detector array (each detector with a 2048$\times$2048 pixel size), and a pixel scale of 0.34 
arcsec pix$^{-1}$. The gaps between the detectors are covered with a series of vertical and horizontal shifts, described in detail 
by \citet{saito12}, but basically, six pawprints with a sky coverage of 0.59 square degrees, are combined into a single tile with a 
1.64 square degrees field of view.

We obtained the photometry for our analysis using the VVV-SkZ pipeline \citep{mauro13}, an automated software based on ALLFRAME 
\citep{stetson94}, and optimized for VISTA PSF photometry. We measured the $J$ and $H$ photometry over the stacked images, 
observed on 27 March, 2010, each with an exposure time of 40 seconds and downloaded from the VISTA Science Archive (VSA) website. 
The $K_S$ photometry was calculated directly from the stacked images, observed between 6 June, 2010 and 1 July, 2013 (36 images). 
Photometric errors are lower than 0.2 mag for $K_S<18$ mag, and for saturated bright stars ($K_S<9.5$ mag) we used 2MASS 
photometry. 

We also spectroscopically observed five stars that are marked with circles in Figure \ref{vvvCL086_Ks} with a single slit in the spectral range 
$2.03-2.34$ $\mu m$ with a resolution of R$\sim$3000, using ISAAC at the Very Large Telescope at ESO Paranal Observatory, Chile 
(on 5 May, 2011). The stars were observed using an ABBA observing mode to later subtract the atmospheric OH emission lines. For 
telluric standards we observed bright A0\,V stars with similar airmass as the cluster stars.

 For spectroscopic data reduction (flat-fielding, sky subtraction, spectra extraction, and wavelength calibration) we used the Interactive 
 Data Language (IDL) and {\sc iraf}\footnote{{\sc iraf} is distributed by the National Optical Astronomy Observatories, which are operated by the 
 Association of Universities for Research in Astronomy, Inc., under cooperative agreement with the National Science Foundation.} scripts, 
 following a similar procedure to that described by \citet{chene12}. For the wavelength calibration we used the OH line, because of sudden 
 shifts along the spectral dispersion direction during the observing night.
 

\begin{table*}
\caption{Individual extinction and distance estimates for stars with spectral classification. Equatorial coordinates, near-infrared magnitudes ($J$, 
$H$, and $K_S$), and spectral classifications are given for all stars (except star \#01, see Section \ref{resultados}). All reported errors are at the
1-$\sigma$ level.}
\begin{center}
\scalebox{0.89}{
\begin{tabular}{ccccccccc}
\toprule
 ID & RA (J2000) & dec (J2000) & $J$ & $H$ & $K_S$ & ST & $A_{K}$ & Distance \\
      & [deg]            &  [deg]             & [mag] & [mag] & [mag]  &    &   [mag]   & [kpc] \\
 \midrule
01 & 252.06556 & -45.43283  & 12.05$\pm$0.01 & 10.68$\pm$0.01 & 10.08$\pm$0.00(3) & $\cdots$ & $\cdots$ & $\cdots$ \\
02 & 252.06546  & -45.43363 & 12.58$\pm$0.01 & 11.28$\pm$0.02 & 10.67$\pm$0.00(3) & B2--3\,V     & 1.33$\pm$0.03 & 1.7$^{+1.7}_{-0.4}$ \\
03 & 252.06539  & -45.43422  &  15.10$\pm$0.02 & 13.78$\pm$0.02 & 13.19$\pm$0.00(3) & O9--B0\,V & 1.38$^{+0.02}_{-0.04}$ & 10.4$^{+3.2}_{-5.0}$ \\
04 & 252.06527  & -45.43570  &  15.59$\pm$0.02 & 14.22$\pm$0.01 & 13.54$\pm$0.00(3) & O9\,V & 1.48$^{+0.01}_{-0.10}$ & 11.7$^{+3.4}_{-3.3}$   \\
05 & 252.06481  & -45.43973  &  18.31$\pm$0.03 & 15.15$\pm$0.01 & 13.83$\pm$0.00(3) & B0\,V &  3.06$\pm$0.03 &  6.2$^{+1.1}_{-2.9}$ \\
\bottomrule
\end{tabular}}
\end{center}
\label{data_stars}
\end{table*}

\section{Results}\label{resultados}

\begin{figure}
\centering
\includegraphics[width=4.2cm,angle=0]{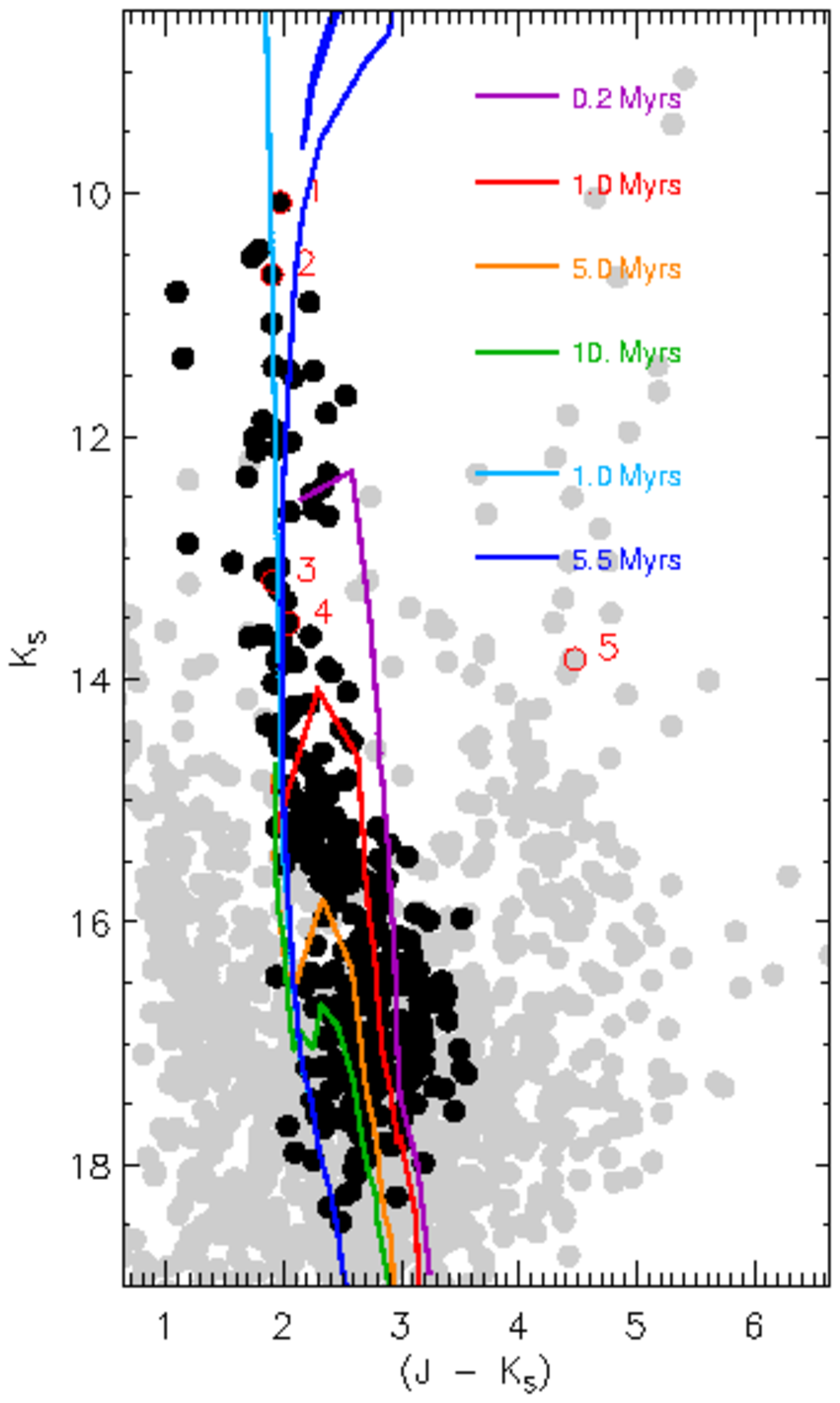}
\hspace{0.1cm}
\includegraphics[width=4.2cm,angle=0]{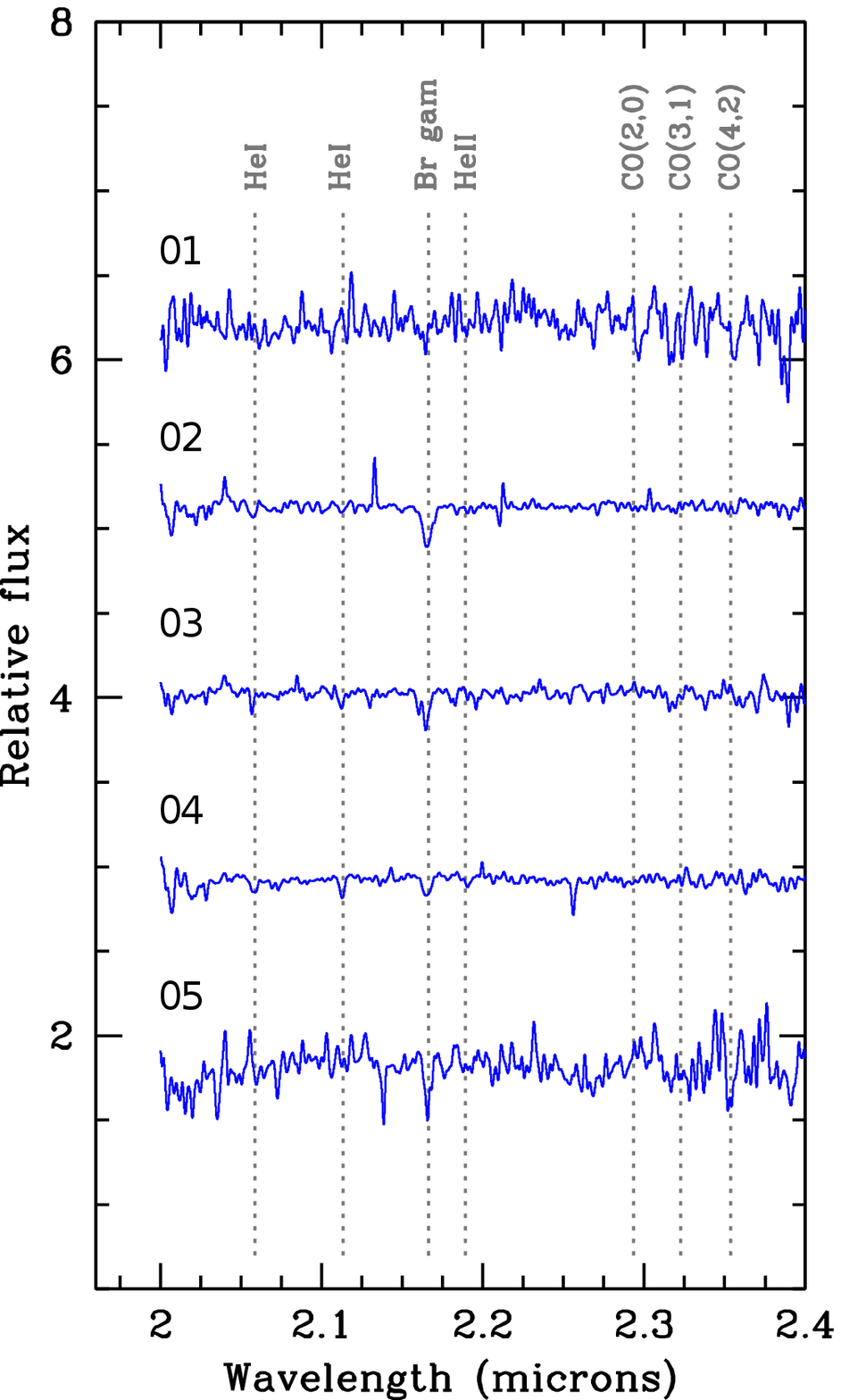}
       \caption{\textit{Left:} VVV\,CL086 field-star decontaminated CMD. The most probable cluster members and field stars are shown with black 
       and grey symbols. Numbers indicate the position for the spectroscopically observed stars. Main (1.0 and 5.5 Myr, 
       \citealt{lejeune01}) and pre-main sequence (0.2, 1.0, 5.0, and 10 Myr, \citealt{siess00}) isochrones are also shown in the 
       diagram. \textit{Right:} individual spectra in $K$ band for VVV\,CL086. We label the spectral lines used for the 
       spectral classification.}
       \label{spectra086}
\end{figure}

In Figure \ref{spectra086} we present the $(J-K_S)$ vs. $K_S$ colour-magnitude diagram. The diagram is statistically field-star decontaminated for a 
circle of radius 0.75 arcmin, centered around ${\alpha}_{2000}=252\fdg0651$, ${\delta}_{2000}=-45\fdg4346$. The decontamination 
was performed as described by \citet{borissova11}, using the algorithm of \citet{bonatto10}. The algorithm divides the $K_S$, $(H-K_S)$ and 
$(J-K_S)$ ranges into a grid of cells with sizes $\Delta K_S$=1.0 mag, and $\Delta (J-K_S)$= $\Delta(H-K_S)$=0.2 mag. In each cell, it 
estimates the expected number density of cluster stars by subtracting the respective field-star number density and, summing over all cells, 
it obtains a total number of member stars, $N_{mem}$. Grid shifts of $\pm$1/3 the cell size are applied in each axes, producing 729 independent 
setups and $N_{mem}$. The average of these 729 $N_{mem}$ (or $\langle N_{mem}\rangle$) is the limit for considering a star as a 
possible cluster member. Only the $\langle N_{mem}\rangle$ with highest survival frequency after all tests were finally considered as cluster members. 
To ensure photometric quality, the algorithm rejects stars with uncertainties in $K_S$ and colours larger than 0.2 mag. For comparison fields we  
used two concentric rings, centred on VVV\,CL086, with inner radius of 0.8$\arcmin$ and 2.5$\arcmin$, and outer radius of 1.2$\arcmin$ and 5.0$\arcmin$, and 
two circles with radius 1.4$\arcmin$: one centred on ${\alpha}_{2000}=252\fdg024$, ${\delta}_{2000}=-45\fdg48$ and the second centred on 
 ${\alpha}_{2000}=252\fdg110$, ${\delta}_{2000}=-45\fdg37$. The stellar densities were corrected for the area difference. The shape and size of 
the control field were chosen based on the distribution of stars and extinction clouds on the image.

In the field-decontaminated diagram we see that stars \#01, \#02, \#03 and \#04 are probably cluster members. Stars \#03 and \#04 are located in 
the cluster main sequence, but would not be the most massive stars in VVV\,CL086. Observed magnitudes and colours for the cluster stellar population 
appear to be affected by differential extinction, a characteristic commonly found in young massive clusters. From its location star \#05, excluded as possible 
cluster member, is expected to be a foreground and very young star, reddened by its surrounding natal gas.
 
 Spectral classification is based on the detection of absorption or emission lines (Br$\gamma$, He\,I at 2.06, 2.11 $\mu m$, and He\,II at 
 2.19 $\mu m$) and the comparison of their shape and depth with similar-resolution spectral catalogues: \citet{hanson96} ($K$ band) 
 and \citet{hanson98} ($H$ band) for OB-type stars. We assumed an error of $\pm 2$ subtypes for the assigned spectral type, similar 
 to \citet{hanson10} and \citet{negueruela10}. The observed spectra and the lines used for the spectral classification are shown in Figure 
 \ref{spectra086}.

 For spectrum \#01 we do not detect any clear spectral feature. The spectrum is very noisy and probably presents faint ${}^{12}$CO\,(2,0) 
 bands at 2.29 $\mu m$. Even if the position of this star in the CMD is consistent with cluster membership, the faint CO band head indicates
 that star \#01 is a very late type object.

 Star \#02 presents a clear Br$\gamma$ line in absorption and He\,I at 2.06 $\mu m$. Its Br$\gamma$ fits the B2\,V HD19734 and B3\,V HD201254 
Br$\gamma$ lines. We adopted spectral type B2-3V for this star. Spectra \#03 and \#04 both show weak Br$\gamma$  and clear He\,I absorption 
line at 2.11 $\mu m$. For star \#03, the lines fit the O9\,V HD193322 and B0.5\,V HD36960 spectra. For star \#04 we also detected the He\,I line 
at 2.06 $\mu m$, and He\,II line at 2.19 $\mu m$, indicating that this star is of earlier type than star \#03. Spectrum \#04 fits O8.5\,V HD73882, and 
O9\,V HD193322 lines. For star \#03 we adopted a spectral type between O9 and B0\,V and for star \#04, spectral type O9\,V. Finally the spectrum 
of star \#05 presents an absorption Br$\gamma$ line that fits a B0\,V spectrum.


\section{Discussion}\label{discusion}

\subsection{Extinction, distance, and radial velocity}

 For individual distance estimates we compared the apparent magnitude with the intrinsic magnitude corresponding to
each individual spectral type. We adopted the \citet{rieke89} extinction law, with $R=3.09$ \citep{rieke85}, 
and intrinsic magnitudes and colours from \citet{martins05}. For stars later than O9.5\,V, we used the intrinsic magnitudes and 
colours from \citet{cox00}. Distance errors are dominated by the spectral type uncertainty, and we estimated it by deriving the 
individual distance for the same star assuming $\pm2$ spectral subtypes.
 
  In Table \ref{data_stars} we present the individual extinction and distance determinations for the spectroscopically observed stars.
Although the position of star \#02 in the CMD associates it with the cluster population, its individual distance estimate indicates that it is a 
foreground star. Using the average of individual distances for stars \#03 and \#04 and the error propagation method described by 
\citet{barlow04}, we obtain a cluster distance of $11^{+5}_{-6}$ kpc. This would locate the cluster in the same region as Mercer 81, 
a massive cluster found at $11\pm2$ kpc \citep{davies12}. The mean extinction value for VVV\,CL086, $A_K=1.5^{+0.0(3)}_{-0.1}$ is lower than 
the extinction value of $A_{2.22}=2.5\pm0.5$ reported for Mercer 81 and indicates the possible presence of an extinction window in that galactic 
direction.

We measured the radial velocities with the {\sc iraf} tasks {\sc fxcor} and {\sc rvidlines} for all the stars with spectral type determination. 
We used the He\,I at 2.076 and 2.11 $\mu m$ and the Br$\gamma$ lines for the estimates. In all cases we measured negative RV, which is expected 
for the cluster direction, but the dispersion due to the number of used spectral lines and the signal-to-noise ratio from our spectra does not allow 
us to obtain a reliable estimate for stellar radial velocities.
 
\subsection{Mass and age estimates}

To estimate the total cluster mass, we first constructed the cluster present-day mass function using the CMD and then integrated
the Kroupa \citep{kroupa01} initial mass function (IMF) fitted to the cluster IMF. Because we did not detect evolved stars, we assumed 
that the present-day and initial mass functions are equivalent.

We obtained the cluster present-day mass function by projecting all CMD stars, following the reddening vector, to the main sequence 
located at 11 kpc. The main sequence is defined by the colours and magnitudes given by \citet{cox00}. After deriving the cluster 
present-day luminosity function, using 1 $K_S$-mag bins, we converted the $K_S$ magnitudes to solar masses using values from 
\citet{martins05} for O-type stars and from \citet{cox00} for stars later than O9.5\,V. 

\begin{figure}
\centering
\includegraphics[width=6.5cm,angle=0]{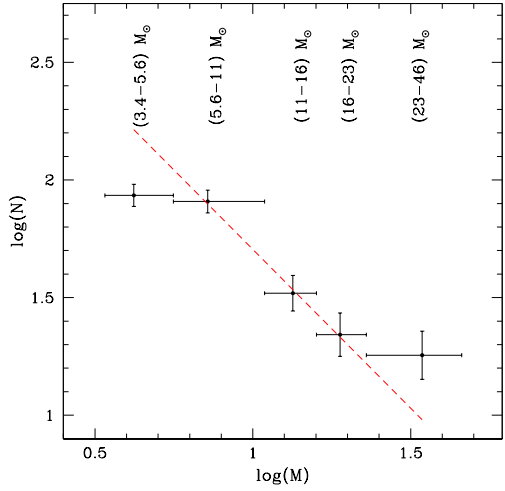}
       \caption{VVV\,CL086 present-day mass function. The points show the central position in the mass ranges 
            indicated above them, and the segmented red line corresponds to the Kroupa IMF fitted to the data. 
            Bar sizes indicate the mass bin equivalent to each magnitude bin (from the luminosity function)
            of 1 mag in $K_S$.}
       \label{vvvCL086_hist}
\end{figure}

The present-day mass function, shown in Figure \ref{vvvCL086_hist}, is fitted by a Kroupa \citep{kroupa01} IMF and 
integrated between 0.10 (log $(M) = -1.00$ dex) and 35 ${M}_{\odot}$ (log $(M) = 1.54$ dex). For the cluster VVV\,CL086 we  
obtain a total mass of $(2.8^{+1.6}_{-1.4})\cdot10^3 {M}_{\odot}$. In our analysis we only included errors associated 
to the fitting of the Kroupa IMF to the data. In the cluster mass function we can see that the cluster could contain a more massive 
population than that spectroscopically detected by us (i.e. stars earlier than O9-B0\,V), but the errors are large and 
dominated by small number statistics. Future spectroscopic observations will help to characterize this population and 
improve the mass estimate.

We tried to determine the cluster age by fitting main sequence \citep{lejeune01} and pre-main sequence \citep{siess00} isochrones. 
However, young (i.e. younger than 10 Myr) cluster main sequences in the infrared are basically vertical lines, which prevented us from
 deriving a precise age for VVV\,CL086. Nonetheless, we observe the pre-main sequence turn-on point in the CMD. Pre-main
 sequence isochrones fitting indicates that the cluster age is $>1.0$ and $<5.0$ Myr.


\section{Conclusions}\label{conclusiones}

We presented the physical characterization of VVV\,CL086, a new massive cluster discovered using data from the VVV survey, found 
at the far edge of the Milky Way bar at a distance of $11^{+5}_{-6}$ kpc. This cluster is the second one found in that 
region of the Galaxy (the first is Mercer 81), a region highly reddened by gas and dust, which presents a relatively low 
mean reddening of $A_K=1.5^{+0.0(3)}_{-0.1}$ mag, however. 

Our spectroscopic follow-up aimed at the brightest stars in the cluster area revealed that two objects are part of the disk population 
(two early-B dwarfs), and two stars form part of the cluster main-sequence population. From their spectral classification and the 
cluster CMD we were able to deduce that earlier stars than these two observed OB-stars are probably be present in VVV\,CL086. One star from our 
spectroscopic follow-up was not classified.

 The mass estimate was derived by integrating the Kroupa IMF fitted to our data and gives a lower limit 
for the cluster total mass of  $(2.8^{+1.6}_{-1.4})\cdot10^3 {M}_{\odot}$. We also estimated the cluster 
age by fitting isochrones to the pre-main sequence turn-on point. We estimated a cluster age 
$>1.0$ and $<5.0$ Myr. The upper age limit agrees with the earliest main sequence star found in the 
cluster (i.e. O9\,V star \#04). Future spectroscopic observations are planned to confirm this and to investigate 
the cluster massive population in more detail.
 
 
\begin{acknowledgements}

We thank to the anonymous referee for the useful comments and suggestions, which helped us to improve this Letter.
 S.R.A. and A.N.C. were supported by the GEMINI-CONICYT project number 32110005. S.R.A. was also supported by 
 the FONDECYT project No. 3140605. The VVV Survey is supported by ESO, by BASAL Center for Astrophysics and Associated 
 Technologies PFB-06, by FONDAP Center for Astrophysics 15010003, and by the Ministry for the Economy, Development, and 
 Tourism's Programa Inicativa Cient\'{i}fica Milenio through grant IC\,12009, awarded to The Millennium Institute of Astrophysics (MAS). 
 Support for J.B. is provided by FONDECYT Regular No.1120601, P.A. acknowledges the support by ALMA-CONICYT project number 
31110002 and D.M., from FONDECYT No. 1130196. M.G. is financed by the GEMINI-CONICYT Fund, allocated to 
Project 32110014. D.G. gratefully acknowledges support from the Chilean BASAL Centro de Excelencia en Astrof\'isica 
y Tecnolog\'ias Afines (CATA) grant PFB-06/2007. R.K.S. acknowledges partial support from FONDECYT through grant 
1130140, and CNPq/Brazil through projects 310636/2013-2 and 481468/2013-7. M.C. acknowledges additional support from 
FONDECYT through grants \#1110326 and 1141141. The work of I.N. and A.M. is partially supported by the Spanish Ministerio de 
Econom\'ia y Competitividad (Mineco) under grant AYA2012-39364-C02-02.

This publication makes use of data products from the Two Micron All Sky Survey, which is a joint project of the 
University of Massachusetts and the Infrared Processing and Analysis Center/California Institute of Technology, 
funded by the National Aeronautics and Space Administration and the National Science Foundation.

\end{acknowledgements}

\bibpunct{(}{)}{;}{a}{}{,} 
\bibliographystyle{aa} 
\bibliography{biblio}

\listofobjects

\end{document}